\documentclass[twocolumn,showpacs,showkeys,preprintnumbers,amsmath,amssymb]{revtex4}

\usepackage{graphicx}
\usepackage{dcolumn}
\usepackage{bm}

\begin{document}


\title{Synchronization of Coupled Anizochronous Auto-Oscillating Systems}

\author{A.~P.~Kuznetsov}%
\affiliation{%
Saratov Branch of the Institute of Radio-Engineering and Electronics of RAS, Zelenaya 38, Saratov 410019, Russia}%

\author{Ju.~P.~Roman}%
 \email{yuliaro@mail.ru}
\affiliation{%
Department of Nonlinear Processes, Saratov State University, Astrakhanskaya 83, Saratov 410012, Russia}%

\date{\today}

\begin{abstract}
The particular properties of synchronization are discussed for coupled auto-oscillating systems, which are characterized by non-quadratic law of potential dependence on the coordinate. In particular, structure of the parameter plane (frequency mismatch -- coupling value) is considered for coupled van der Pol -- Duffing oscillators. The arrangement of synchronization tongues and the particular properties of their internal structure in the parameter space are revealed. The features of attractors in the phase space are discussed.
\end{abstract}

\maketitle

\section{\label{sec:level1}Introduction}

The system of two coupled van der Pol oscillators is one of the basic models of nonlinear dynamics demonstra\-ting the phenomenon of mutual synchronization. There are many papers on this theme [see 1 - 15] because this system demonstrates a lot of interesting oscillation regimes and types of behavior, such as synchronous and quasiperiodic regimes, synchronization with different phase conditions [1, 15], the "oscillation death" effect [1, 2], global bifurcations [7], etc. Moreover, the results of this system investigation may be used in the analysis of different electronic, biological and chemical systems [3 - 5, 11 - 13]. Interest in this problem does not decrease because of its significance and dynamics variety [see 14].

For the van der Pol system is only single non-linearity characteristic. It is non-linear dissipation, which limits amplitude of auto-oscillation. Potential function is the same as in the case of the harmonic oscillator. Deve\-loping this system, we consider van der Pol -- Duffing oscillator, for which additional non-linearity of Duffing oscillator type is introduced [1, 15]. Potential contains in this case a component in the form of the coordinate raised to the fourth power. This component takes into account an important physical effect, namely an opportunity of anisochronous oscillations, i.e. the dependence of oscillation frequency on oscillation amplitude [1, 15].

One of the most important methods of such systems synchronization analysis is the quasiharmonic approximation. It consists in the construction of equations for the amplitude and the phase, which are slowly changing against the background of oscillations with the eigenfrequency of oscillators [1, 15]. But this method is effective only when the excess above the threshold of bifurcation of auto-oscillation appearance is small. So it designates only the structure of the main synchronization tongue. At the same time the overall picture of the parameter space structure for coupled auto-oscillating systems is substantially not yet to be established. (It is in contrast to the case of forced synchronization by external harmonic signal [1, 16, 17].) It concerns especially the case of anisochronous system, when the structure of characte\-ristic regime areas depends on the additional parameter, which is responsible for the form of potential function.

In the present paper the parameter space structure is investigated for coupled auto-oscillating systems with several characteristic types of potential function, which are leading to the anisochronous oscillations. There are such cases considered, when quasiharmonic approximation is ineffective or unusable at all.

\section{Coupled van der Pol oscillators}

The system of differential equations describing the interaction between auto-oscillating systems is of the form
\begin{equation}
\begin{split}
&\frac{\displaystyle d^{2}x}{\displaystyle dt^{2}}-(\lambda-x^{2})\frac{\displaystyle dx}{\displaystyle dt}+\frac{\displaystyle \partial U_{1}(x)}{\displaystyle \partial x}+\mu(\frac{\displaystyle dx}{\displaystyle dt}-\frac{\displaystyle dy}{\displaystyle\displaystyle dt})=0,\\
&\frac{\displaystyle d^{2}y}{\displaystyle dt^{2}}-(\lambda-y^{2})\frac{\displaystyle dy}{\displaystyle dt}+\frac{\displaystyle \partial U_{2}(y)}{\displaystyle \partial y}+\mu(\frac{\displaystyle dy}{\displaystyle dt}-\frac{\displaystyle dx}{\displaystyle dt})=0.\\
\end{split}
\end{equation}
Here $\lambda$ is the parameter characterizing the excess above the threshold of Andronov -- Hopf bifurcation in \makebox{autonomous} oscillators, $\mu$ is the coefficient of dissipative coupling, $U_{1}(x)$ and $U_{2}(y)$ are the potential functions of each oscillator. Their difference determines non-identity of oscillators.

\begin{figure*}
\includegraphics{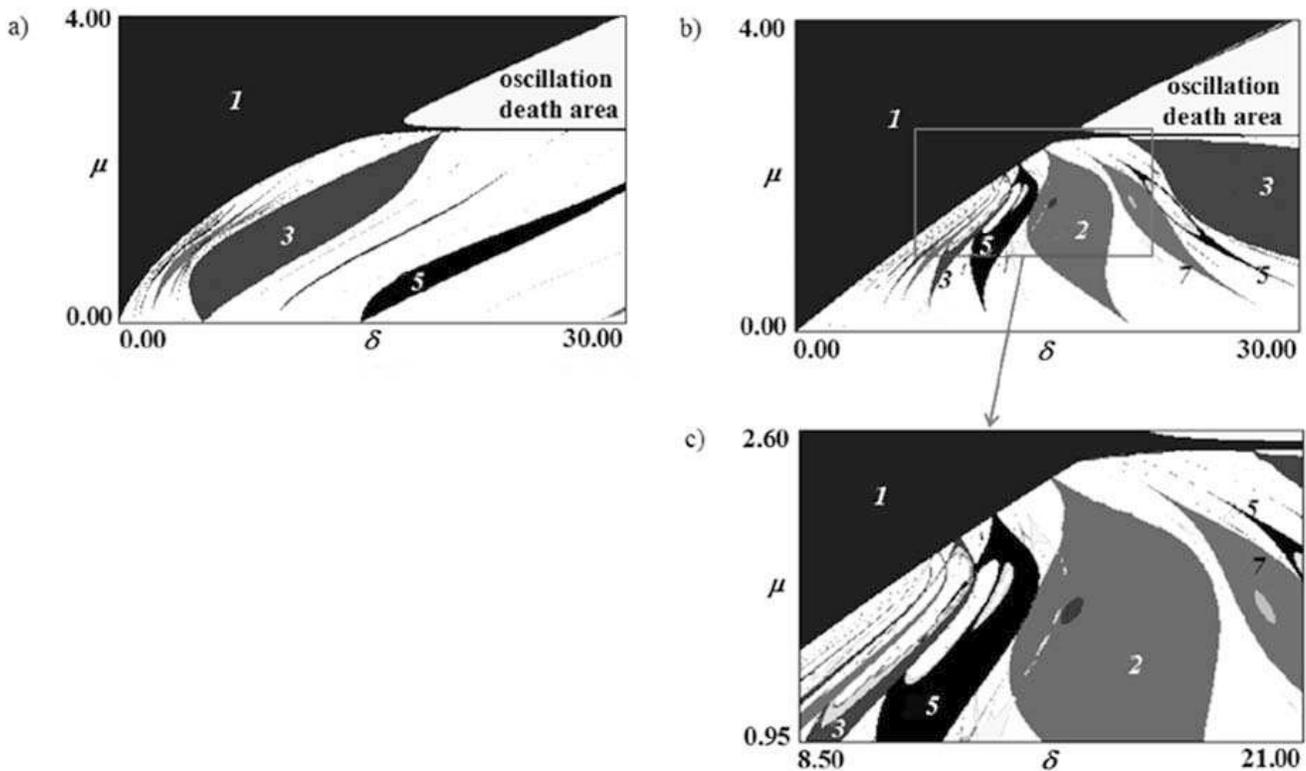}
\caption{\label{fig1:wide} Dynamic regime charts (a)~for the system (3) for $\lambda=2.5$ and (b,~c)~for the system (5) for $\lambda=2.5$, $\beta=1$. Numbers on some synchronization tongues denote cycles periods.}
\end{figure*}
In simple case of quadratic potential
\begin{equation}
U_{1}(x)=\frac{x^2}{2},\quad U_{2}(y)=(1+\delta)\frac{y^2}{2}
\end{equation}
the system (1) is the system of two coupled van der Pol oscillators, where parameter $\delta$ is the frequency mismatch between the second and the first oscillators:
\begin{equation}
\begin{split}
&\ddot{x}-(\lambda-x^{2})\dot{x}+x+\mu(\dot{x}-\dot{y})=0,\\
&\ddot{y}-(\lambda-y^{2})\dot{y}+(1+\delta)y+\mu(\dot{y}-\dot{x})=0.\\
\end{split}
\end{equation}

To determine the parameter space structure of the system (3) we shall use the method of dynamic regime chart construction [18]. Within the framework of such a method we shall mark the oscillation period of the system of coupled oscillators by means of different colors (gray color hues) on the parameter plane (frequency mismatch $\delta$ -- coupling value $\mu$). White color corresponds to the chaotic or quasiperiodic motions. Cycle periods were calculated by means of the Poincare section method: this is the number of points of intersection of the phase trajectory on the attractor and the surface chosen as the Poincare section. Only those crossings were taken into account that correspond to the trajectories coming to the surface from the one side. The number of points of intersection of the phase trajectory and the intersecting surface was considered as the oscillation period.

The system under investigation is characterized by four-dimensional phase space $(x,\dot{x},y,\dot{y})$. Therefore three-dimensional hypersurface that is preset by means of some additional condition, e.g. zero velocity of the se\-cond oscillator $\dot{y}=0$ may serve as the Poincare section. In that case the number $n$ of points of intersection of trajectory and the hypersurface was determined. Colors on the charts are chosen in accordance with the period $n$.

The chart of dynamic regimes obtained in such a way for the van der Pol oscillators (3) is given in Fig.~\ref{fig1:wide} (a) on the ($\delta$, $\mu$)-plane for $\lambda=2.5$. The value of the control parameter is chosen so large, that traditional quasiharmonic approximation [1, 14, 15] is not efficient.

On the chart in Fig.~\ref{fig1:wide} (a) there are shown: main synchronization tongue with the period 1; area of quasipe\-riodic regimes with the embedded system of higher order synchronization tongues; the area of the "oscillation death" effect [1, 2] which corresponds to stabilization of the equilibrium state point at the origin due to sufficiently strong dissipative coupling.

\section{Coupled van der Pol -- Duffing oscillators}

Now we pass on to the van der Pol -- Duffing model, to which the potentials
\begin{equation}
U_{1}(x)=\frac{x^2}{2}+\beta \frac{x^4}{4},\quad U_{2}(y)=(1+\delta)\frac{y^2}{2}+\beta \frac{y^4}{4}
\end{equation}
and dynamic equations in the form of
\begin{equation}
\begin{split}
&\ddot{x}-(\lambda-x^{2})\dot{x}+x+\beta x^{3}+\mu(\dot{x}-\dot{y})=0,\\
&\ddot{y}-(\lambda-y^{2})\dot{y}+(1+\delta)y+\beta y^{3}+\mu(\dot{y}-\dot{x})=0.\\
\end{split}
\end{equation}
correspond.

The parameter $\beta$ defines perturbation quantity of the quadratic potential and, correspondingly, is responsible for anisochronism of small oscillations [1, 15].

On the charts in Fig.~\ref{fig1:wide} (b, c) there are shown the dynamic regime chart for the system (5) and its enlarged fragment for $\beta=1$ that is the case of sufficiently great anisochronism. One can see in Fig.~\ref{fig1:wide} (b) that the pre\-sence of anisochronism leads to the displacement of synchronization tongues towards the greater values of frequency mismatch. Synchronization tongues become so wide that one can see the situation of their overlapping, which is characteristic for standard circle map [18]. Internal structure of this tongues changes too: period \makebox{doublings} and transition to chaotic dynamics inside them are observed. It may be seen clearly in enlarged fragment of the dynamic regime chart, shown in Fig.~\ref{fig1:wide} (c).

\begin{figure}[h!]
\includegraphics[scale=0.6]{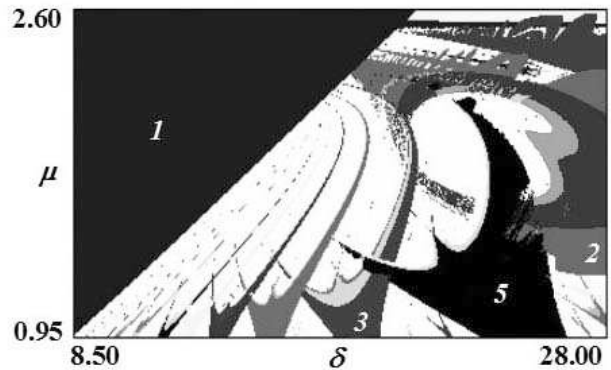}
\caption{\label{fig2:epsart} Dynamic regime chart for the system (5) for $\lambda=2.5$, $\beta=3$.}
\end{figure}

With the increase of the parameter of anisochronism $\beta$ an onset of new "islands" of doubled periods takes place inside the synchronization tongues locating at greater values of frequency mismatch $\delta$ (see Fig.~\ref{fig2:epsart}). Synchronization tongues change their shape with the further increase of parameter $\beta$. They fall into two similar in structure pieces, inside which structures called "crossroad area" exist [18]. These structures are characteristic for systems with period doublings. Practically, two systems of synchronization tongues may be observed in the parameters plane: tops of the first one are disposed along the line of zero coupling, tops of the second one -- along the boun\-dary of main synchronization area 1, which is the line of Neimark -- Sacker bifurcation. Chaotic area appears between these two systems of synchronization tongues.

\begin{figure*}
\includegraphics[scale=0.35]{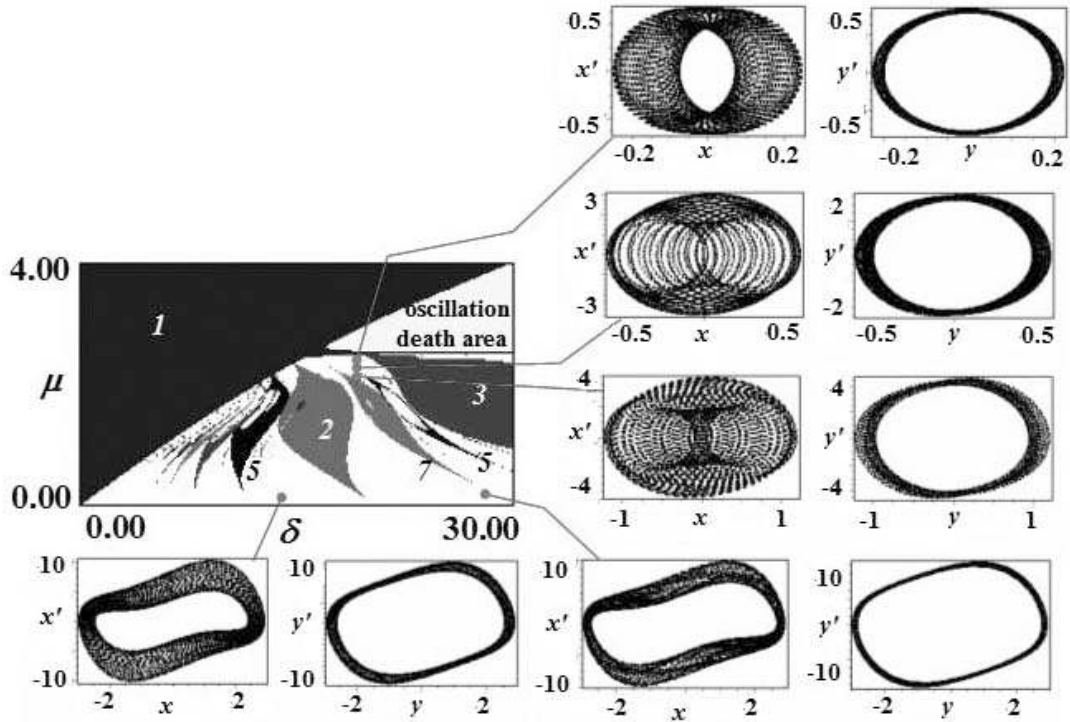}
\caption{\label{fig3:wide} Phase plane portraits computed at characteristic points within the area of quasiperiodic regimes for the system (5) for $\lambda=2.5$, $\beta=1$.}
\end{figure*}

Phase plane portraits of attractors computed at se\-veral characteristic points within the domains of quasiperiodic regimes are given in Fig.~\ref{fig3:wide} on $(x,\dot{x})$ and $(y,\dot{y})$ planes for the first and the second oscillators correspondingly. To put it more precisely these portraits are projections of the attractor in phase space $(x,\dot{x},y,\dot{y})$ onto correspon\-ding planes $(x,\dot{x})$ and $(y,\dot{y})$. For small values of coupling parameter it may be seen that portraits of attractors look like slightly disturbed limit cycles of individual oscillators. At the same time the trajectory never approaches the neighborhood of the point of origin. With the increase of the coupling parameter the trajectories of individual oscillators become more perturbed and there exists a kind of a threshold value when the trajectory of the first oscillator may achieve the neighborhood of the origin. Corresponding phase portrait looks like the area completely "filled" with trajectories. Evidently, this si\-tuation corresponds to the case of ill-determined phase of this oscillator. When approaching the oscillation death area an essential downsizing of attractors occurs (see scales in coordinate axis). The orbit ceases again to achieve neighborhood of the origin, but structure of phase portraits and shapes of attractors are different from the case of small coupling value.

\begin{figure*}
\includegraphics[scale=0.9]{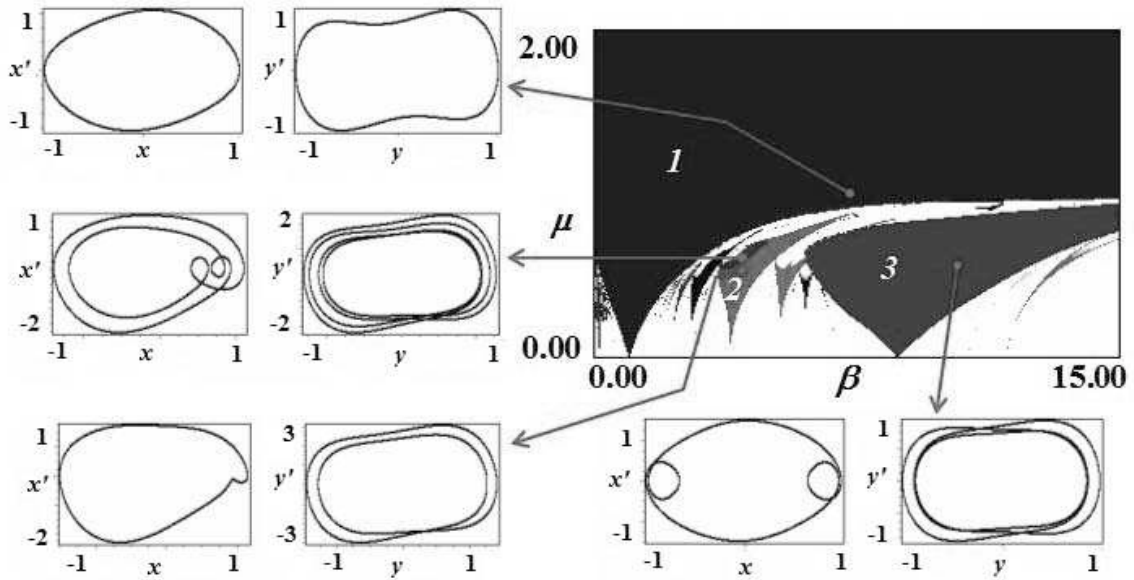}
\caption{\label{fig4:wide} Dynamic regime chart for the system (7) for $\lambda=1$ and phase plane portraits computed at several chosen points within the synchronization tongues with periods 1, 2, 3 and 4.}
\end{figure*}

\section{Coupled oscillators with the potential of the fourth degree}
Now we consider the case when the potential does not contain quadratic component and is in the form of
\begin{equation}
U_{1}(x)=\beta_{1} \frac{x^4}{4},\quad U_{2}(y)=\beta_{2} \frac{y^4}{4}.
\end{equation}
Corresponding dynamic equations are
\begin{equation}
\begin{split}
&\ddot{x}-(\lambda-x^{2})\dot{x}+\beta_{1} x^{3}+\mu(\dot{x}-\dot{y})=0,\\
&\ddot{y}-(\lambda-y^{2})\dot{y}+\beta_{2} y^{3}+\mu(\dot{y}-\dot{x})=0.\\
\end{split}
\end{equation}

We shall note that the potential of this type was offered by Ueda for common (i.e. not auto-oscillating) oscillators. Corresponding model was called Ueda's oscillator [18].

In this case it is impossible to talk about eigenfrequencies of oscillators and, therefore, to introduce their mutual frequency mismatch. But we shall provide for non-identity of oscillators by means of different values of the parameter $\beta$, which is responsible for the shape of the potential well (6). If we choose $\beta_{1}=1$ for the first oscillator, we shall vary the parameter $\beta_{2}=\beta$.

Structure of the parameter plane $(\beta,\mu)$ constructed for the system (7) is shown in Fig.~\ref{fig4:wide} for $\lambda=1$. It is remarkable that the system of synchronization tongues with different periods exists as before in spite of the fact that slightly excited autonomous oscillators have no fixed eigenfrequencies of small oscillations at all. Nevertheless, different steepness of the potential well of the fourth degree results in possibility of different rhythms in separate oscillators and in synchronization with various periods.

Absence of the oscillation death area in spite of dissipative nature of coupling is yet another essentially new moment. This fact may be easily proved by means of combined equations (7). In order to this regime was possible, the point of origin $x=y=0$ should be attracting. System of equations (7) linearized about this point has the form of
\begin{equation}
\begin{split}
&\ddot{x}+(\mu-\lambda)\dot{x}=\mu \dot{y},\\
&\ddot{y}+(\mu-\lambda)\dot{y}=\mu \dot{x}.\\
\end{split}
\end{equation}

Looking for the solution of the system (8) in the form of $e^{\alpha t}$ , we shall obtain simple characteristic equation
\begin{equation}
\alpha^{2}(\alpha-\lambda)(\alpha+2\mu-\lambda)=0.
\end{equation}

One can see that at least one root of this equation $\alpha=\lambda$ is always positive when auto-oscillations exist and \makebox{$\lambda>0$}. So state of equilibrium in the point of origin is always unstable. At the same time the condition $\mu=\lambda$ becomes apparent in the structure of the dynamic regime chart. It is the boundary of the main synchronization tongue when $\beta\gg1$ and oscillators (7) are strongly non-identical. One more feature consists in the fact that tops of the multiple synchronization tongues are disposed along the lower boundary of the main synchronization tongue, i.e. their structure differs from the case shown in Fig.~\ref{fig1:wide} (b) in the area of great values of coupling parameter.

Phase portraits of the first and the second oscillators computed inside the basic synchronization tongues are given in Fig.~\ref{fig4:wide}. Regime of doubled period on the basis of synchronous regime 2 is also shown in this figure.
\begin{figure}
\includegraphics[scale=0.3]{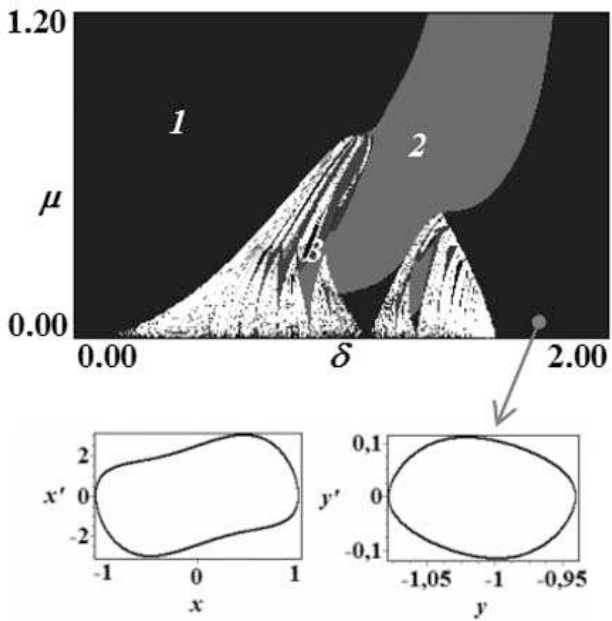}
\caption{\label{fig5:epsart} Dynamic regime chart and phase plane portrait computed at characteristic point for the system with the potential of the fourth degree (10) for $\lambda=2.5$.}
\end{figure}

Non-identity of oscillators in the case of potential of the fourth degree may be introduced in one more way, namely by means of mutual displacement of the potential wells of interacting oscillators:
\begin{equation}
U_{1}(x)=\frac{x^4}{4},\quad U_{2}(y)=\frac{(y+\delta)^4}{4}.
\end{equation}

Parameter $\delta$ controls in this case the position of the minimum of the fourth degree for the second oscillator. The dynamic regime chart is shown for this system in Fig.~\ref{fig5:epsart}. We also observe the system of synchronization tongues. An interesting feature of synchronization picture consists in this case in the fact that with the \makebox{excess} above the value $\delta\approx1$ synchronization with the period 1 displaces all other regimes and occurs even if the \makebox{coupling} parameter is indefinitely small. Phase plane portrait given in Fig.~\ref{fig2:epsart} indicates the fact that the first oscillator dominates over the second oscillator (see scales in coordinate axis).

\section{Conclusion}

The analysis of synchronization in auto-oscillating systems by means of the dynamic regime chart construction reveals an essential role of the anisochronism. If we insert nonlinearity of Duffing oscillator type, synchronization tongues become essentially wider and the situation of their overlapping occurs. Islands of doubled periods, "crossroad area" structures and chaotic areas appear inside the synchronization tongues. For oscillators with the potential of the fourth degree we have observed the system of synchronization tongues, which depends on the parameter that controls relative steepness of the potential wells of oscillators. The oscillation death effect disappears in this case and synchronization tongues modify themselves in the area of great values of coupling parameter. If mutual position of the potential wells of the fourth degree is changing, there is a critical value when synchronization with the period 1 occurs even if the \makebox{coupling} parameter is indefinitely small.

\section*{Acknowledgements}

This work was supported by the grant RFBR (Project No. 06-02-16773) and the foundation of nonprofit programs "Dynasty".

\end{document}